\documentclass[10pt,twocolumn]{article}
\usepackage{nopageno}
\usepackage{times}

\usepackage{authblk}
\usepackage[dvipsnames]{xcolor}
\usepackage{color}
\definecolor{darkgray}{gray}{0.25}
\usepackage[normalem]{ulem}
\usepackage{placeins}
\usepackage[hyphenbreaks]{breakurl}

\usepackage[export]{adjustbox}

\usepackage[subtle]{savetrees}
\makeatletter
\renewcommand\paragraph{\@startsection{paragraph}{4}{\z@}%
                                      {0.25ex \@plus1ex \@minus.2ex}%
                                      {-1em}%
                                      {\normalfont\normalsize\bfseries}}
\makeatother

\usepackage[hyphens]{url}
\usepackage[hidelinks]{hyperref}
\hypersetup{
	colorlinks,
	citecolor=Maroon,
	linkcolor=Maroon,
	urlcolor=blue,
	breaklinks=true
}
\usepackage{cite}

\usepackage{xspace}
\usepackage{paralist}
\usepackage{graphicx}
\usepackage{verbatim}
\usepackage{comment}
\usepackage{fixltx2e}
\usepackage{float}
\usepackage{lipsum}

\graphicspath{ {figs/} }

\title{A Secure Cloud with Minimal Provider Trust
\footnote{{\scriptsize DISTRIBUTION STATEMENT A. Approved for public release: distribution unlimited.
This material is based upon work supported by the Assistant Secretary of Defense for Research
and Engineering under Air Force Contract No. FA8721-05-C-0002 and/or FA8702-15-D-0001. Any
opinions, findings, conclusions or recommendations expressed in this material are those of
the author(s) and do not necessarily reflect the views of the Assistant Secretary of Defense
for Research and Engineering. Delivered to the U.S. Government with Unlimited Rights, as
defined in DFARS Part 252.227-7013 or 7014 (Feb 2014). Notwithstanding any copyright notice,
U.S. Government rights in this work are defined by DFARS 252.227-7013 or DFARS 252.227-7014
as detailed above. Use of this work other than as specifically authorized by the U.S.
Government may violate any copyrights that exist in this work.}}
}
\author{Amin Mosayyebzadeh$^{1}$, Gerardo Ravago$^{1}$, Apoorve Mohan$^{6}$, Ali Raza$^{1}$, Sahil Tikale$^{1}$,

 Nabil Schear$^{2}$, Trammell Hudson$^{3}$, Jason Hennessey$^{1,4}$,

 Naved Ansari$^{1}$, Kyle Hogan$^{5}$, Charles Munson$^{2}$,

Larry Rudolph$^{3}$, Gene Cooperman$^{6}$, Peter Desnoyers$^{6}$, Orran Krieger$^{1}$
\\
\small {\em  $^1$Boston University, Boston, MA \quad
  $^2$MIT Lincoln Laboratory, Lexington, MA  \quad
  $^{3}$Two Sigma, New York, NY \quad
  $^{4}$NetApp Inc.\quad
\\  $^{5}$MIT, Cambridge, MA \quad
  $^{6}$Northeastern University, Boston, MA} \vskip 2mm
}

\date{}

\begin{document}

\maketitle

\begin{abstract}
	Bolted is a new architecture for a bare metal cloud with the goal of providing security-sensitive customers of a cloud the same level of security and control that they can obtain in their own private data centers.  It allows tenants to elastically allocate secure resources within a cloud while being protected from other previous, current, and future tenants of the cloud. The provisioning of a new server to a tenant isolates a bare metal server, only allowing it to communicate with other tenant's servers once its critical firmware and software have been attested to the tenant.  Tenants, rather than the provider, control the tradeoffs between security, price, and performance.  A prototype demonstrates scalable end-to-end security with small overhead compared to a less secure alternative.

\end{abstract}

\section{Introduction}

Despite all the advantages of today's public clouds, many security sensitive organizations are reluctant to use them because of their security challenges and the trust that the tenant needs to place in the cloud provider. Can we make a cloud that is appropriate for even the most security sensitive tenants? Can we make a cloud where the tenant does not need to fully trust the provider? Can we do this without hurting the performance of tenants that do not wish to pay for extra security?

The key security challenge of Infrastructure-as-a-Service (IaaS) clouds stems from collocating multiple tenants on a single physical node with virtualization. Malicious tenants can exploit vulnerabilities in the huge trusted computing base (TCB) to launch attacks on tenants running on the same node~\cite{king2006subvirt} or even worse, if the attacker compromises the hypervisor, to launch attacks on the cloud provider. Moreover, virtualization enables side-channel and covert channel attacks such as the recent Meltdown and Spectre exploits~\cite{ristenpart2009hey,Kocher2018spectre,Lipp2018meltdown,liu,razavi2016flip}.  
Recent processor secure enclave technology Intel SGX has suffered from its own security challenges from the collocation of multiple tenants~\cite{branch-shadowing,controlled-channel,BMDKCS17,Sanctum,sgxspectre}. 
Such security concerns keep huge sections of the economy, such as medical companies and hospitals, financial institutions, federal agencies etc., from being able to take advantage of the benefits of today's clouds.~\cite{bucci_getting_2012, paquette_identifying_2010, govt_IT_report_2017, 4_reasons_2013}\footnote{ IARPA recently released RFI~\cite{IARPA_rfi_2017} describing the requirements of one security-sensitive community to {\em replicating as closely as possible the properties of an air-gapped private enclave}. We believe that meeting this requirement would alleviate the concerns of a broad community of security-sensitive customers, making the geographical distribution, elasticity, and on-demand pricing of cloud available to a wide community of users.}

Bare metal clouds~\cite{Softlayer:2015,RackSpace:2015,Internap:2015,Packet.net,AWSBare} remove the threat of side-channel attacks and covert channels implicit in virtualization. However, all the existing bare metal clouds still require the tenant to fully trust the provider which may not always be a safe assumption. Consider how one protects against server firmware attacks. If a prior tenant of a node is able to inject malicious firmware, this modified firmware can be used to attack future tenants of that server. Existing clouds protect against firmware attacks on the tenant's behalf~\cite{AWSBare, IntelTrusted} but there is no way for the tenant to verify, for example, that the provider was not compromised or even has installed all firmware security patches. 

Bolted differs from today's bare metal clouds by reducing the implicit trust in the provider. Bolted allows a tenant to elastically carve out a secure private enclave of commodity physical servers in which she may run applications. The enclave is protected from previous users of the same servers (using hardware-based attestation), concurrent tenants of the cloud (using network isolation and encryption), and future users of the same servers (using storage encryption and memory scrubbing). With Bolted an organization with security expertise is able to deploy their own attestation infrastructure, and can directly validate the measurements against their expectations of firmware and software deployed. Each Bolted component addresses a different potential vulnerability. It is the combination of Bolted components that minimizes the role of the provider to mostly that of securing the physical access to the hardware.

We have implemented a prototype of Bolted and demonstrate that the prototype can provision an enclave of sixteen servers with a full application environment with legacy servers and firmware in around 8 minutes, enabling highly elastic environments. We show that the overhead to do attestation with Bolted is low, adding only 25\% to the cost of a highly optimized provisioning system. Replacing traditional UEFI firmware with a customized Linux-based firmware we developed for this service, we further improve the security of the user and reduce provisioning time to being around 10\% faster than the unattested version; a node can be fully provisioned with a full application environment in just over 3 minutes.

\vspace{-12pt}
\section{Threat Model}\label{threat}
 
Our goal in Bolted is to enable tenants to strongly isolate themselves from other tenants while placing as little trust in the provider as possible.  
Specifically, we trust the provider to maintain the physical security of the hardware, so physical attacks like bus snooping or de-capping chips are out of scope.  
We also trust the provider for availability of the network and node allocation services and any network performance guarantees.  
We assume that all cloud provider nodes are equipped with Trusted Platform Modules (TPMs)~\cite{TPM_trusted_2008}.


We categorize the threats that the tenant faces into the following phases:

\paragraph{Prior to occupancy:} Malicious (or buggy) firmware can threaten the integrity of the secure enclave of which the node becomes a part. We must ensure that a previous tenant (e.g., by exploiting firmware bugs) or cloud provider insider (e.g., by unauthorized firmware modification) did not infect the node's firmware prior to the tenant receiving it. Further, we must ensure that the node being booted is isolated from potential attackers until it is fully provisioned and all defenses are in place. 

\paragraph{During occupancy:} Although many side-channel attacks are avoided by disallowing concurrent tenants on the same server, we must ensure that the node's network traffic is isolated so that the provider or other concurrent tenants of the cloud cannot launch attacks against it or eavesdrop on its communication with other nodes in the enclave. Moreover, if network attached storage is used (as in our implementation) all communication between storage and the node must be secured.

\paragraph{After occupancy:} We must ensure that the confidentiality of a tenant is not compromised by any of its state (e.g, storage or memory) being visible to subsequent software running on the node.

\vspace{-12pt}
\section{Design Philosophy}

A central design principle of Bolted is to enable as much functionality as possible to be implemented by the tenant rather than by the provider for three reasons: (i) to minimize the trust that a tenant needs to place in the provider, (ii) to enable tenants with specialized security expertise and requirements to implement functionality themselves, and (iii) to enable tenants to make their own cost/performance/security tradeoffs. This principle has a number of implications for our design.

First, Bolted differs from existing bare metal offerings in that most of the component services that make up Bolted can be operated by a tenant rather than by the provider. A security sensitive tenant can customize or replace these services. All the logic, that orchestrates how different services are used to securely deploy a tenant's software, is implemented using scripts that can be replaced or modified by the user.

Second, while we expect a provider to secure and isolate the network and storage of tenants, we only rely on the provider for availability and not for the confidentiality or integrity of the tenant's computation. In the most secure deployments, we assume that Bolted tenants will further encrypt all communication between the tenants' nodes and between those nodes and storage. Bolted provides a (user-operated) service to securely distribute keys for this purpose.

Third, we rely on attestation (measuring all firmware and software and ensuring that it matches known good values) that can be implemented by the tenant rather than just validation (ensuring that software/firmware is signed by a trusted party). This is critical for firmware which may contain bugs~\cite{thunderstrike,thunderstrike2,lighteater,atr-smm,x86-harmful,heasman} that can disrupt tenant security. Attestation provides a time-of-use proof that the provider has kept the firmware up to date. More generally, we attest through the process of incorporating a node into an enclave, and we can also continuously attest when the node is operating, to ensure that bugs in any layer of software (irrespective of who signed them) have not allowed malicious code to be executed.

Fourth, we have a strong focus on keeping our software as small as possible and making it all available via open source. In some cases, we have written our own highly specialized functionality rather than relying on larger function rich general purpose code in order to achieve this goal. For functionality deployed by the provider, this is critical to enable it to be inspected by tenants to ensure that any requirements are met. For example, previous attacks have shown that firmware security features are difficult to implement bug-free -- including firmware measurements being insufficient~\cite{chronomancy}, hardware protections against malicious devices not being in place~\cite{iommu-attack}, and dynamic root of trust (DRTM) implementation flaws~\cite{wojtczuk2009attacking}.
Further, our firmware is deterministically built, so that the tenant can not only inspect it for correct implementation but then attest that this is the firmware that is actually executing on the machine assigned to the tenant. For tenant deployed functionality, small open source implementations are valuable to enable user-specific customization.

\vspace{-12pt}
\section{Architecture}

\begin{figure}
	\includegraphics[width=8cm]{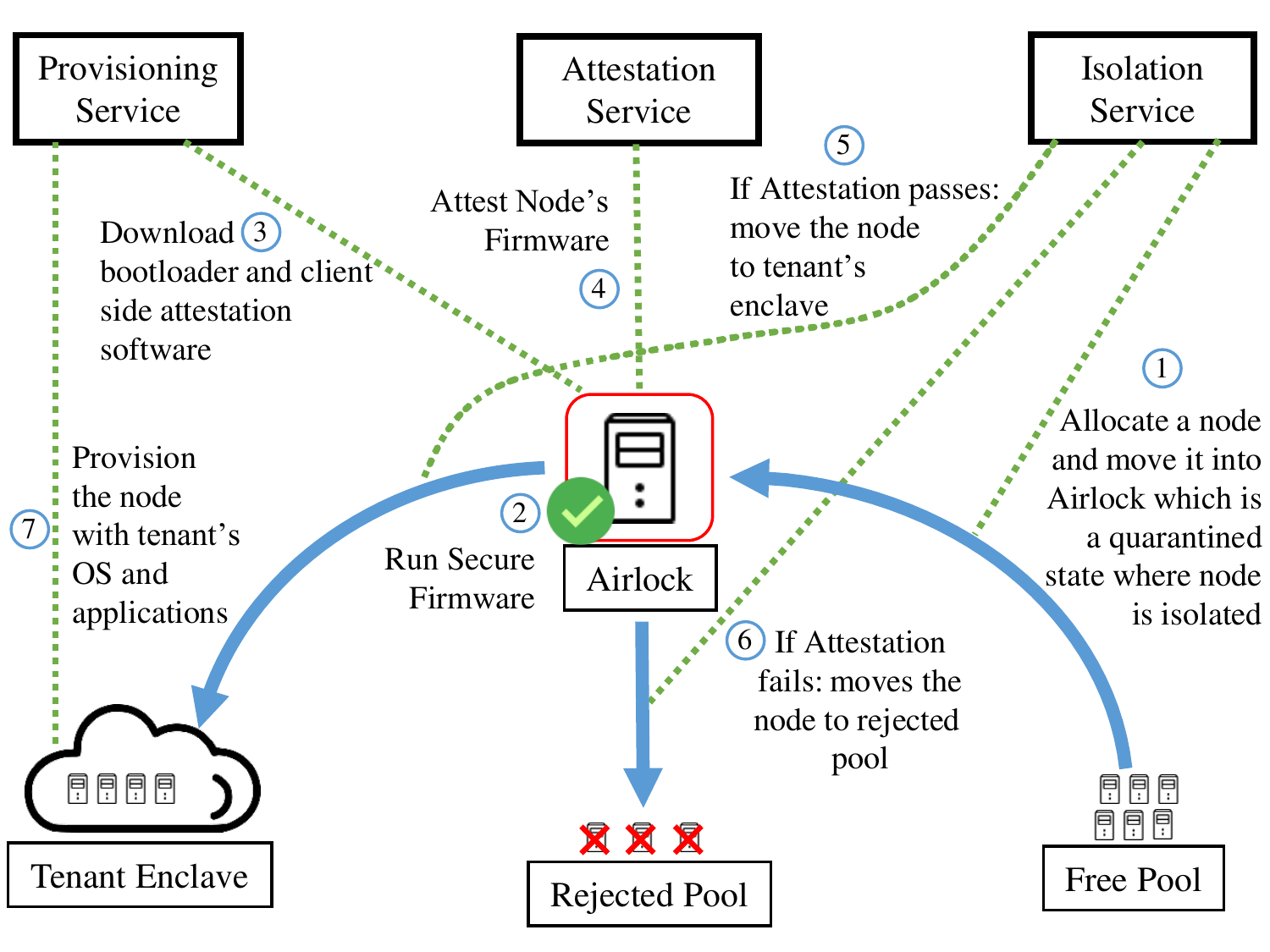}
        \vspace{-0.25cm}
	\caption {Bolted's architecture: Blue arrows show state changes and green dotted lines show actions taken by each service.}
	\label{fig:simple}
        \vspace{-0.25cm}
\end{figure}

The architecture of Bolted and the sequence that a node goes through as part of being admitted to a tenant enclave is shown in Figure~\ref{fig:simple}. The state of the node changes as a result of its interaction with the three main services that comprise the Bolted system. Like any bare metal offering, Bolted requires, an {\em isolation service} which allocates nodes and configures networks to isolate those nodes from nodes of other tenants and a {\em provisioning service} to provision the user's operating system and applications onto the allocated nodes. The Bolted architecture adds a third: an {\em Attestation Service}.

The attestation service consists of a server and a client component that runs in the firmware of the node.  The server is responsible for: 1) maintaining a whitelist of trusted firmware/software measurements, 2)  comparing  {\em quotes} (hash measurements signed by the TPM) of firmware/software against the whitelist, 3) maintaining a {\em registry} of TPM to node mappings to verify quotes by the TPM and ensure that the quotes are coming from an expected node in the cloud, and 4) distributing keys to nodes so that they may encrypt communication between them as well as securely accessing storage.  

The client component is responsible for participating the node in the attestation protocol. On boot, the hardware measures the first portion of firmware that in turn measures the remainder of the firmware and the next code (e.g., a bootloader) before it is executed. That code in turn loads, measures and then executes subsequent software, etc. These measurements are all stored in the TPM. 
The client obtains quotes from the TPM, i.e., cryptographically signed measurements of the firmware and software that are loaded and executed on the node. The client securely provides the quotes to the attestation server, which then matches them to its whitelist. The compromised software will not match the whitelist and the infected node will be rejected. Upon successful attestation, the verifier securely provides the node a cryptographic key that can bootstrap encrypted storage and network isolation. At this point, the node is acquired by the tenant. It is critical that client firmware and software be open, simple, verifiable, and, ideally, deterministically built to enable reliable and secure attestation. 

\vspace{-12pt}
\section{Implementation}\label{impl_sec}

We have developed a prototype of the Bolted architecture. Here, we briefly describe the implementation components of the three main services, the node firmware we use and how these components work together.

\paragraph{Hardware Isolation Layer:}
The Hardware Isolation Layer (HIL)~\cite{hil} is our implementation of the Bolted isolation service. The fundamental operations HIL provides are (i) allocation of physical nodes, (ii) allocation of networks, and (iii) connecting these nodes and networks. A tenant can invoke HIL to allocate nodes to an enclave, create a management network between the nodes, and then connect this network to any provisioning tool (e.g.,~\cite{emulab, ironic, bmi, maas}). She can then create additional networks for isolated communication between nodes and/or attach those nodes to public networks made available by the provider. HIL is a very simple service (approximately 3000 LOC). It creates networks (currently VLANs~\cite{vlan}) and attaches nodes to them by interacting with the switches of the provider.

\paragraph{Malleable Metal as a Service:}
The Malleable Metal-as-a-Service (M2)~\cite{bmi} is our implementation of the Bolted provisioning service.
The fundamental operations M2 provides are: (i) create (disk) image, (ii) clone and snapshot an image, (iii) delete an image, and (iv) boot a node from an image. Similar to virtualized cloud services, M2 services images from remote-mounted boot drives. Images are exposed to the nodes via an iSCSI (TGT~\cite{tgt}) service managed by M2 and stored in a Ceph~\cite{ceph} distributed file system. As published previously, M2 is between 3-4 times faster than traditional provisioning systems that install an image into a server's local disk~\cite{bmi}.

\paragraph{Keylime:}
Keylime\cite{keylime} is our implementation of the Bolted attestation service. It is divided into four components: Registrar, Cloud Verifier, Client and Tenant. The registrar maintains the node to TPM mapping. The verifier maintains the whitelist of trusted code and checks nodes' integrity. The Keylime client is downloaded and measured by the node firmware and then passes quotes from the node's TPM to the verifier. Keylime Tenant starts the attestation process and asks Verifier to verify the node which runs Keylime client.

\paragraph{Heads:}
Heads~\cite{heads} is our firmware implementation and bootloader replacement. It is a minimal deterministically built version of Linux that (i) zeroes all node memory, (ii) downloads the Keylime client, and (if attestation has succeeded) (iii) downloads and kexecs to a tenant's kernel. In the future, we expect to directly mount the iSCSI disk from M2 to obtain the kernel, but currently, we fetch the kernel from a web service stood up for this purpose.

\paragraph{Putting it together:}
The booting of a node is controlled by a Python application that follows the sequence of steps shown in Figure~\ref{fig:simple}. Secure OS images contain a Keylime client that obtains a key to encrypt network traffic between nodes in the enclave as well as traffic to the boot disk mounted using iSCSI.
For servers that support it, we burn Heads directly into the server's flash, and for the other servers, we download Heads from a PXE service stood up for this purpose and then continue the same sequence as if Heads was burned into the flash. We have modified iPXE client code to measure the downloaded Heads image into a TPM register so that all software involved in booting a node can be attested.



\vspace{-12pt}

\section{Evaluation}\label{eval_sec}

We show performance results from our initial prototype implementation of Bolted. Timing breakdowns are shown using a Dell R630 server with 256 GB RAM and 2 2.6GHz 10 (20 HT) core Intel Xeon processors model E5-2660 v3. We have physical access to this server in our lab and show experiments with both stock UEFI firmware and our own Heads firmware. Scalability experiments are shown on 16 Dell M620 blade serves with 64 GB memory and 2 2.60GHz 8 (16 HT) core Xeon E5-2650 v2 processors connected to a 10Gbit switch.


\begin{figure}
	\includegraphics[width=8cm,left]{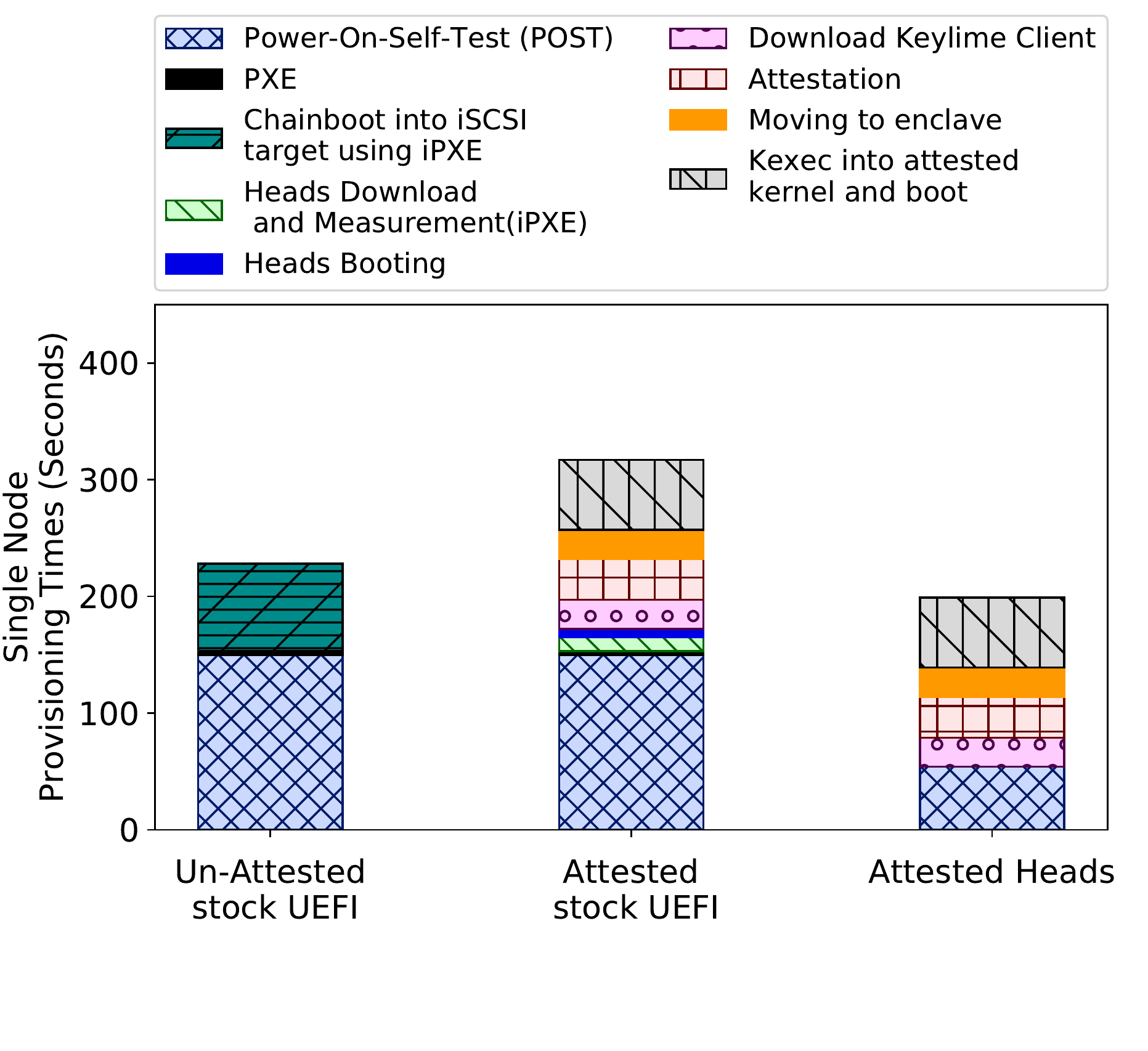}
        \vspace{-1.5cm}
	\caption {Performance with and without Bolted for systems with stock UEFI firmware.}
	\label{fig:detailed_stack_kumo}
        \vspace{-0.25cm}
\end{figure}

Figure~\ref{fig:detailed_stack_kumo} shows the timing breakdown of different stages of provisioning a 20 GB image of pre-installed linux with Bolted. The three scenarios are: 1) an unattested boot that just uses M2 to directly boot a users image, 2) fully attested boot with stock UEFI firmware, and 3) fully attested boot with Heads burnt into the flash. 

Without Bolted's security features, a server node provisioning via M2 takes three steps: Power-on-self-test (POST), PXE, and then chainbooting from an iSCSI target using iPXE. The total time is under 4 minutes with 2.5 minutes spent in the POST step alone.

For full attestation using stock UEFI firmware, after POST, Bolted goes through the following phases: (i) PXE downloading iPXE, (ii) iPXE downloading and measuring Heads, (iii) booting Heads, (iv) download the Keylime client (currently using http) and measuring it, (v) running the Keylime client, registering the node and attesting it, and then downloading (currently using http) and measuring the tenants kernel, (vi) moving the node into the tenants enclave and, and finally (vii) Heads kexec to the tenants kernel and it is booted\footnote{This implementation is a very early prototype, and we expect to be able to speed up steps 4 and 5 by incorporating iSCSI drivers into our Heads implementation.}. With all these steps the total time to provision a server is just over 5 minutes or around 25\% more than the unattested boot. 

For full attestation using Heads firmware, after POST we immediately jump to step 4 above. Heads posts in just under a minute (almost half of which on this server is a timeout waiting for Intel Management Engine to initialize). With Heads burned into the firmware total provisioning time is 35\% faster than the fully attested case with stock firmware and even 10\% faster than the unattested case with stock firmware.




\begin{figure}[ht]
        \vspace{-0.25cm}
	\includegraphics[width=8cm, left]{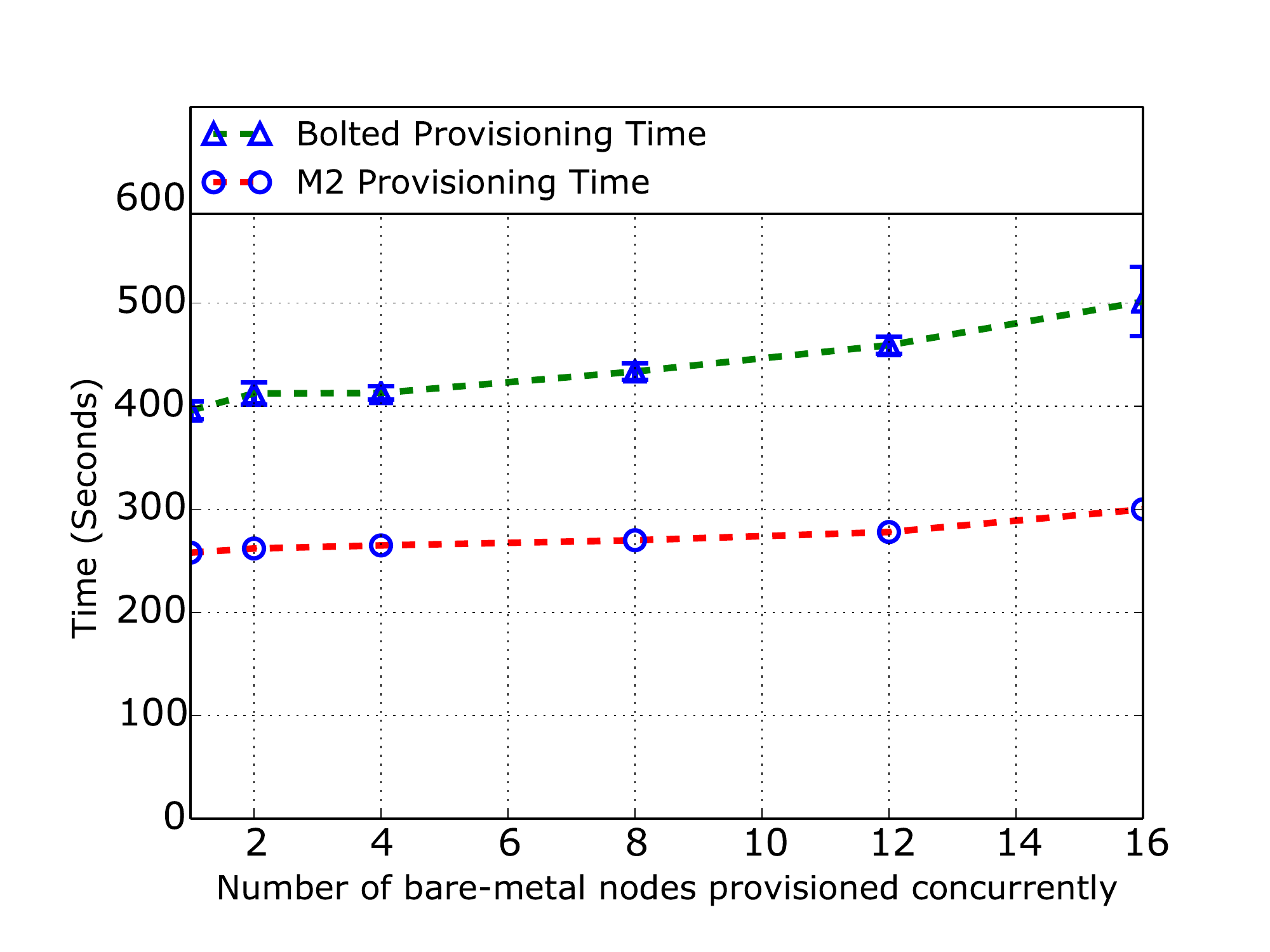}
        \vspace{-0.75cm}
	\caption {Scaling Results.}
	\label{fig:scaling}
        \vspace{-0.25cm}
\end{figure}

Figure~\ref{fig:scaling} shows (with UEFI firmware) the scalability of Bolted with and without attestation as we increase the number of concurrently booting nodes. Each experiment was run five times, and the line shows the degradation in performance for the average of those runs as we increase the number of nodes from one to sixteen. With sixteen servers booting concurrently, performance degrades by around 50 seconds for the 
unattested case and around 100 seconds for the attested case. Degradation in the unattested experiments is due to a large number of concurrent block requests to the small scale Ceph deployment (with only 27 disks). Experiments~\cite{bmi} with a larger scale Ceph deployment demonstrated better M2 scalability. Performance of a fully attested boot degrades by only 13\% as we move to 16 servers.



\vspace{-12pt}
\section{Discussion}
In this paper, we describe Bolted, an architecture for a bare metal cloud that is appropriate for even the most security sensitive tenants. For these tenants, Bolted enables protection from attacks: (i) prior to occupancy using attestation to ensure that any nodes with compromised firmware are rejected and by isolating nodes (in airlocks) until they can be added to tenant enclaves, (ii) during occupancy by allocating entire servers to avoid virtualization attacks and by providing a secure model to distribute tenant keys for encrypting storage and network traffic, (iii) after occupancy by using firmware that scrubs memory prior to booting other software and using M2 for network mounted storage. The only trust these tenants need to place in the provider is: (a) the availability of the resources and (b) that the physical hardware has not been compromised.

The only Bolted service that needs to be deployed by a provider is the isolation service (e.g., HIL), which needs to be trusted by the provider to control the physical switches. All other services can be deployed by a tenant or on their behalf by a third party and the orchestration to enable an attested boot is managed by scripts controlled by the tenant. This means that a security sensitive tenant can operate these services in their own environment. Customers with specialized needs may choose to develop their own variants of these services.

The cost/complexity/performance/security trade-offs are fully under the tenants control. A tenant that doesn't want the cost and complexity to deploy their own instance of Keylime and M2, or that wants to take advantage of a large-scale implementation by the provider, can choose to trust provider-deployed versions of these services. Also, if a tenant chooses to trust the network isolation of the provider (e.g. HIL) he/she may feel no need to encrypt network and/or storage traffic. Finally, tenants that are willing to trust firmware validation (e.g. firmware is bug-free and signed by the vendor) are free to do that and will not incur any of the performance overhead of attestation. 

To enable a wide community to inspect them and minimize their TCB, all components of Bolted are open source, including Keylime~\cite{github_python_keylime_2018}, Heads~\cite{hudson_heads_2018}, M2~\cite{github_malleable_2018}, and HIL~\cite{github_hil_2018}.
We designed HIL, for example, to be a simple micro-service rather than a general purpose tool like IRONIC~\cite{ironic} or Emulab~\cite{emulab}.
HIL is being incorporated into a variety of different use cases by adding tools and services on and around it rather than turning it into a general purpose tool. Another key example of a small open source component is Heads. Heads is much simpler than UEFI.
Since it is based on Linux, it has a code base that is under constant examination by a huge community of developers.
Heads is reproducibly built, so a tenant can examine the software to ensure that it meets their security requirements and then ensure that the firmware deployed on machines is the version that they require.
For example, the firmware must measure all of itself before launching the next level of software.
As another example, we need to make sure that firmware zeros all the memory of a server before enabling subsequent software to run\footnote{We need to zero memory in the firmware since a malicious provider may steal a server from a tenant at any point and we need to make sure that all state is removed from the node before the next tenant can see it.}.

We are not able to flash Heads on all servers, and the servers we have flashed it on require physical access~\cite{heads}. However, we are working with the OpenCompute community to both enable Heads to be flashed remotely and to ensure that OpenCompute vendors provide Heads as a supported option; it is enormously difficult without vendor support to ensure that servers with minor changes will successfully boot. If we cannot flash our own firmware, Bolted uses stock firmware to download Heads. In this situation, Heads provides us with a standardized execution environment for the Keylime client and download the tenant kernel. While we have no guarantee that the stock firmware is up-to-date and fully measured, in this situation Bolted provides attestation against the white list of the most up-to-date firmware to ensure known vulnerabilities have been addressed.  

Bolted protects against compromise of firmware executable by the system CPU; however modern systems may have other processors with persistent firmware inaccessible to the main CPU; compromise of this firmware is not addressed by this approach. 
These include: Base Management Controllers (BMCs)~\cite{moore_2017}, the Intel Management Engine~\cite{newman_2017,ermolov_goryachy_2017,kroizer_2015}, PCIe devices with persistent flash-based firmware, like some GPUs and NICs, and storage devices~\cite{hd_virus}.
Additional work (e.g. techniques like IOMMU use, disabling the Management Engine~\cite{me_cleaner} and the use of systems without unnecessary firmware) may be needed to meet these threats.

While it is a work in progress, our early scalability results are encouraging, even in an initial prototype on a testbed with several known performance and scalability issues.
They suggest that a complete implementation of Bolted with Heads burned into the firmware and a larger scale storage backend will enable us to elastically provision dozens, perhaps even hundreds of fully attested servers in under five minutes.
If we can achieve this, it will make Bolted appropriate for highly-elastic security-sensitive situations, e.g., a national emergency requiring many computers.
This will hugely reduce the need for institutions that keep around large numbers of largely idle machines to deal with low probability events.

\vspace{-12pt}

\section{Acknowledgment}
We would like thank Intel, Red Hat, Two Sigma, NetApp and Cisco,  the core industry partners of MOC for supporting this work.  We gratefully acknowledge Piyanai Saowarattitada and Radoslav Nikiforov Milanov for their significant contributions in development and their assistance in the evaluations. 
Partial support for this work was provided by the USAF Cloud Analysis Model Prototype project, National Science Foundation awards CNS-1414119, ACI-1440788 and OAC-1740218.

\clearpage

{\footnotesize \bibliographystyle{acm}
\bibliography{refs,db,bolted}}

\end{document}